\documentclass[english,twocolumn,aps,pra,showpacs]{revtex4}
\usepackage[T1]{fontenc}
\usepackage[latin1]{inputenc}
\usepackage{graphicx}
\usepackage{amssymb}

\makeatletter

\input{epsf}
\makeatother

\usepackage{babel}
\makeatother

\begin{document}

\title{Finite temperature phase diagram of a spin-polarized ultracold Fermi gas in
a highly elongated harmonic trap}

\author{Xia-Ji Liu${^1}$, Hui Hu$^{1,2}$, and Peter D. Drummond${^1}$}

\affiliation{$^{1}$\ ARC Centre of Excellence for Quantum-Atom Optics, Department
of Physics, University of Queensland, Brisbane, Queensland 4072, Australia\\
$^{2}$\ Department of Physics, Renmin University of China, Beijing
100872, China}

\date{\today{}}

\begin{abstract}
We investigate the finite temperature properties of an ultracold atomic
Fermi gas with spin population imbalance in a highly elongated harmonic
trap. Previous studies at zero temperature showed that the gas stays in an
exotic spatially inhomogeneous Fulde-Ferrell-Larkin-Ovchinnikov (FFLO)
superfluid state at the trap center; while moving to the edge, the system
changes into either a non-polarized Bardeen-Cooper-Schriffer superfluid 
($P<P_c$) or a fully polarized normal gas ($P>P_c$), depending on the
smallness of the spin polarization $P$, relative to a critical value $P_c$.
In this work, we show how these two phase-separation phases evolve with
increasing temperature, and thereby construct a finite temperature phase
diagram. For typical interactions, we find that the exotic FFLO phase
survives below one-tenth of Fermi degeneracy temperature, which seems to be
accessible in the current experiment. The density profile, equation of
state, and specific heat of the polarized system have been calculated and
discussed in detail. Our results are useful for the on-going experiment at
Rice University on the search for FFLO states in quasi-one-dimensional
polarized Fermi gases.
\end{abstract}


\maketitle

\section{Introduction}

Impressive experimental progress has occurred recently in the
field of ultracold Fermi gases. In particular, experimental realizations
of the crossover from a Bardeen-Cooper-Schrieffer (BCS) superfluid to
Bose-Einstein condensate (BEC) using a well-controlled Feshbach resonance 
\cite{jila,mit04,duke,innsbruck,ens,rice05,natphys} have opened an
intriguing opportunity to study long-standing many-body problems in
condensed matter physics, such as high temperature superconductivity.
Most recently, two experimental groups at Rice University and MIT have
successfully manipulated a two-component atomic Fermi gas of lithium atoms
with \emph{unequal} spin populations \cite {mit06a,rice06a,rice06b,mit06b,mit06c,mit07}. 
This type of matter is of great interest, and stimulates intense 
efforts on studying an unsolved problem in condensed matter and 
particle physics: what is the ground state of a spin-polarized Fermi 
gas with attractive interactions?

Without spin polarization, the answer is known, given fifty
years ago by Bardeen, Cooper, and Schrieffer. Due to attractions, atoms
of different spin species at the \emph{same} Fermi surface with opposite
momentum form Cooper pairs characteristic of bosons, and thus undergo a
BEC-like superfluid phase transition at a sufficiently low temperature. This
BCS mean-field picture is very robust and is valid not only quantitatively
at weak-coupling, but also qualitatively in the strongly interacting limit 
\cite{natphys}, where the fluctuations of Cooper pairs become important. The
key ingredient of the BCS pairing is the fully overlapped Fermi surfaces,
which create a maximum for the phase space. In the presence of spin
polarization, however, the two Fermi surfaces are no longer aligned. The 
standard Cooper pairing scheme is thus not applicable. For a small number 
of unpaired atoms, or, a small spin polarization, some non-standard pairing 
scenarios have to be developed. For a large spin polarization above threshold,
these unpaired atoms will eventually destroy the coherence of pairs and
hence the superfluidity.

The determination of non-standard pairing scenarios and the related exotic
superfluidity lies at the heart of polarized Fermi gases. A number of
pairing proposals have already been suggested in the literature, including
breached pairing \cite{liu} or Sarma superfluidity \cite{sarma,yip},
phase separation \cite{bedaque}, deformed Fermi surface \cite{muther}, and
spatially modulated Fulde-Ferrell-Larkin-Ovchinnikov (FFLO) states \cite{ff,lo}. 
At the moment, a theoretical consensus on the true ground state of a three-dimensional
spin-polarized Fermi system is yet to be reached \cite{sheehy,huixiajipra,son,lianyihe,iskin,xiajihuiepl,parish}.
Experimental observations near a Feshbach resonance are not so helpful as 
one would expect, due to the presence of the harmonic trap. To take it into 
account, the local density approximation has been widely 
used \cite{chevy,yi,silva,lobo,chien,bulgac,haque}, as well as the
mean-field Bogoliubov-de Gennes equation \cite{kinnunen,machidaprl,xiajipra3d}.

Among various pairing schemes, the study of FFLO phases has the longest
history---more than four decades \cite{ff,lo}. In this
scenario, the two mismatched Fermi surfaces have been shifted by an amount
in momentum space \cite{ff}, in order to have a small overlap of the
surface. As a result, Cooper pairs with a \emph{finite} center-of-mass
momentum may form, and thereby support a spatially inhomogeneous
superfluidity \cite{lo}. However, due to the much reduced phase space for
pairing, the window for the appearance of FFLO states turns out to be small 
in three dimensions \cite{sheehy,huixiajipra}. This makes the experimental 
search for the FFLO states extremely challenging. Till now only indirect 
evidence has been observed in the heavy fermion superconductor CeCoIn$_5$ \cite{cecoin5}.
No FFLO signal has been found in ultracold polarized atomic Fermi gases.

Luckily, the FFLO states are theoretically found to be favorable in low dimensions, 
which can be realized experimentally using an optical lattice \cite{moritz}. 
In one dimension (1D), where the whole Fermi surface shrinks to two Fermi points, 
the reduction of pairing phase space is less significant. Consequently, at zero 
temperature the FFLO phase becomes much more robust \cite{machida,buzdin,yang}, 
compared to its 3D counterpart \cite{huixiajipra}. This expectation was indeed 
found in recent works \cite{guan1,guan2,orso,xiajiprl,xiajipra1d,parish2}. Further,
the presence of a harmonic trap will not change the essential picture, as firstly 
shown by Orso \cite{orso} and the present authors \cite{xiajiprl,xiajipra1d}.
Nonetheless, the trap leads to phase separation \cite{orso,xiajiprl}: While the 
system remains always in the FFLO phase at the trap center, at the trap edge, 
it can be either a fully non-polarized BCS state or a fully polarized normal 
state, for a small or large spin polarization, respectively. Most recently, 
the spatial distribution of pair correlation functions and the momentum distribution
of a spin-polarized 1D Fermi gas in lattices have been investigated by using 
numerically accurate density-matrix renormalization-group methods \cite{feiguin,tezuka,luscher} 
and quantum Monte Carlo simulations \cite{batroun}.

Inspired by theoretical suggestions, an experiment with a polarized
Fermi gases in a highly elongated harmonic trap is now underway at Rice
University \cite{hulet}. However, the experiment will necessarily be carried
out at a finite temperature. It is then desirable to understand how the two
phase separation phases found at zero temperature evolve as temperature
increases, and in particular, what is the temperature window for the
presence of FFLO states, compared to the lowest experimentally attainable
temperature.

In a previous study \cite{xiajipra1d}, we have presented a systematic study 
of zero-temperature quantum phases in a one-dimensional spin-polarized Fermi gas. 
Three theoretical methods have been comparatively used: the mean-field
theory with either an order parameter in a single-plane-wave form or a self-consistently 
determined order parameter using the Bogoliubov-de Gennes equations 
\cite{bdg,reidl,bruun,nygaard,grasso,kinnunen,machidaprl,xiajipra3d}, as well as the exact 
Bethe ansatz method \cite{takahashi}. We have found that a spatially inhomogeneous 
FFLO, which lies between the fully paired BCS state and the fully polarized normal state, 
dominates most of the phase diagram of a uniform gas. The phase transition from the 
BCS state to the FFLO phase is of second order. We have also investigated the effect 
of a harmonic trapping potential on the phase diagram, and find that in this case the trap
generally leads to phase separation as mentioned above. We finally investigate the local 
fermionic density of states of the FFLO phase. A two-energy-gap structure is shown up, 
which may be used as an experimental probe of the FFLO states. 

In this work, we address the urgent finite-temperature problem by extending our previous
zero-temperature analysis. In particular, we focus on the use of weak-coupling Bogoliubov-de 
Gennes (BdG) theory \cite{bdg,bruun,kinnunen}. Based on this mean-field approach we 
are able to obtain the density profiles of the system at finite temperatures, as well 
as its thermodynamic properties, including the entropy, energy and specific heat.

%
\begin{figure}
\begin{centering}\includegraphics[clip,width=0.45\textwidth]{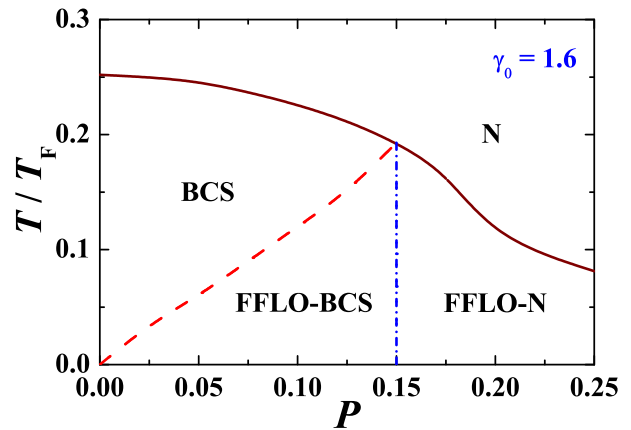}\par\end{centering}

\caption{(color online) Finite temperature phase diagram of a 1D
trapped spin-polarized Fermi gas at a given coupling constant $\gamma _0=1.6$, 
as a function of the spin polarization, 
$P=(N_{\uparrow }-N_{\downarrow})/(N_{\uparrow }+N_{\downarrow })$, 
where $N_{\uparrow }$ and $N_{\downarrow }$ are the number of spin-up and 
spin-down atoms, respectively. $T_F$ is the Fermi temperature of an ideal, 
non-interacting Fermi gas in a harmonic trap. The solid line gives the 
boundary of the second order phase transition from a superfluid to a 
normal state. The dashed and dot-dashed lines distinguish the different 
superfluid phases: a pure BCS phase, a phase separation with a FFLO at 
center and a BCS outside (FFLO-BCS), and a phase separation with FFLO and 
normal components (FFLO-N). Due to the presence of harmonic traps, 
these boundaries may be better understood as crossover, rather than 
phase transition lines.}

\label{fig1} 
\end{figure}


Our main result, a finite temperature phase diagram, is shown in Fig.
1 for a typical interaction strength. At a finite but low temperature, in
addition to the two phase separation states mentioned earlier, two new
phases---a pure BCS state and a partially polarized normal state---enter 
the phase diagram, respectively, at lower and higher spin polarizations. The
space for the phase separation states shrinks with increasing temperature
and vanishes completely at one-fifth of the Fermi temperature $T_F$.
Therefore, even at a temperature as high as $0.1T_F$, it is still possible
to observe the long-sought FFLO state at the center of the harmonic trap.
This suggests that the FFLO state in 1D polarized Fermi gases is indeed
attainable with current experimental techniques \cite{natphys}.

In the following section, we briefly review the mean-field BdG theory of the 1D 
spin-polarized Fermi gas. The applicability of the mean-field theory in the weakly
coupling or intermediate coupling regimes will be commented. In Sec. III, we will 
discuss in detail how the density profiles and the order parameters evolve with
increasing temperatures. Together with the analysis of the entropy, energy and 
specific heat, we obtain a phase diagram at finite temperatures. Finally, Sec. IV 
is devoted to the conclusions and remarks.
 
\section{Self-consistent Bogoliubov-de Gennes theory}

Fermi gases of $^6$Li atoms near a broad Feshbach resonance can be well
described using a single channel model, as confirmed both experimentally \cite{sc1} 
and theoretically \cite{sc2,sc3}. To obtain the phase diagram in Fig. 1, we use 
the self-consistent BdG theory \cite{bdg,reidl,bruun}, by assuming a pairing order 
parameter $\Delta (x)$ that breaks the $U(1)$ symmetry of the number conservation 
of total neutral atoms. This weak
coupling theory has been previously applied to discuss the zero temperature
properties of 1D polarized Fermi gases by the present authors. 
We refer to Ref. \cite{xiajipra1d} for the detailed description of the 
model and the BdG formalism. We outline below some essential 
ingredients of the theory.

The BdG equations describing the quasiparticle wave functions $u_\eta \left(
x\right) $ and $v_\eta \left( r\right) ,$ with excitation energies $E_\eta $
are \cite{bdg}: 
\begin{equation}
\left[ 
\begin{array}{cc}
{\cal H}_{\uparrow }^s-\mu _{\uparrow } & -\Delta (x) \\ 
-\Delta ^{*}(x) & -{\cal H}_{\downarrow }^s+\mu _{\downarrow }
\end{array}
\right] \left[ 
\begin{array}{c}
u_\eta \left( x\right) \\ 
v_\eta \left( x\right)
\end{array}
\right] =E_\eta \left[ 
\begin{array}{c}
u_\eta \left( x\right) \\ 
v_\eta \left( x\right)
\end{array}
\right] ,  \label{BdG}
\end{equation}
where the single particle Hamiltonian ${\cal H}_\sigma ^s=-\hbar ^2\nabla
^2/2m+g_{1D}n_{\overline{\sigma }}\left( x\right) +V_{ext}\left( x\right)$,
and $V_{ext}\left( x\right) =m\omega ^2x^2/2$ is the harmonic trap
potential. The interaction is a short range potential 
$g_{1D}\delta \left( x\right) $, resulting in a diagonal Hatree term 
$g_{1D}n_{\overline{\sigma }}\left( x\right) $ and an off-diagonal order
parameter potential in the mean-field decoupling. To account for the unequal
spin population $N_\sigma $ for the pseudospins 
$\sigma =\uparrow,\downarrow $, the chemical potentials are shifted 
as $\mu _{\uparrow,\downarrow }=\mu \pm \delta \mu $. This leads to 
different quasiparticle wave functions for the two spin components. However, 
there is a symmetry of the BdG equations under the replacement 
$u_{\eta \downarrow }^{*}\left( x\right)\rightarrow v_{\eta \uparrow }\left( x\right)$, 
$v_{\eta \downarrow}^{*}\left( x\right) \rightarrow -$ $u_{\eta \uparrow }\left( x\right)$, 
$E_{\eta \downarrow }\rightarrow -E_{\eta \uparrow }$. Thus, in Eq.(1) we
have kept only the spin-up part, and obtained the solutions with both
positive and negative excitation energies.

The order parameter and the particle density of each component that appears
in the BdG equations can be written as, 
\begin{eqnarray}
\Delta \left( x\right) &=&-g_{1D}\sum_\eta u_\eta (x)v_\eta ^{*}(x)f(E_\eta
),  \label{gap} \\
n_{\uparrow }\left( x\right) &=&\sum_\eta u_\eta ^{*}(x)u_\eta (x)f(E_\eta ),
\label{nup} \\
n_{\downarrow }\left( x\right) &=&\sum_\eta v_\eta ^{*}(x)v_\eta
(x)f(-E_\eta ),  \label{ndw}
\end{eqnarray}
where $f(E_\eta )=1/(\exp [E_\eta /k_BT]+1)$ is the Fermi distribution
function. The particle density of each component is subjected to the
constraint $\int dx$ $n_\sigma \left( x\right) =N_\sigma $, which
eventually determines the chemical potentials. Eq. (\ref{BdG}), together with
the definitions of the order parameter and particle densities, 
Eqs. (\ref{gap}), (\ref{nup}), and (\ref{ndw}), form a closed set of 
equations, which has to be solved self-consistently. We have done so via 
a \emph{hybrid} procedure, in which the high-lying excitation levels above 
an energy cutoff $E_c$ has been solved approximately using a local density 
approximation. This procedure is very efficient. We refer to 
Ref. \cite{xiajipra1d,reidl} for further details. Note that the final 
numerical results are independent of the cutoff energy $E_c$, provided 
that it is large enough.

Once the BdG equations are solved, it is straightforward to calculate the
total entropy and total energy of the system. The expression for the entropy
is given by, 
\begin{equation}
S=-k_B\sum\limits_{E_\eta }\left[ f(E_\eta )\ln f(E_\eta )+f(-E_\eta )\ln
f(-E_\eta )\right] ,  \label{entropy}
\end{equation}
where the summation can be restricted to energy levels with energy 
$\left| E_\eta \right| \leq E_c$. Other high-lying branches have an exponentially 
small contribution because of the large value of $E_c$($\gg k_BT$). This 
has been neglected. In contrast, the total energy includes two 
parts, $E=E_{disc}+E_{cont}$. The discrete contribution $E_{disc}$ takes 
the form, 
\begin{eqnarray}
E_{disc} &=&\left[ \mu _{\uparrow }N_{\uparrow }+\mu _{\downarrow
}N_{\downarrow }-g_{1D}\int dx\left| \Delta (x)\right| ^2\right]  \nonumber
\\
&&+\sum\limits_{\left| E_\eta \right| \leq E_c}E_\eta \left[ f\left( E_\eta
\right) -\int dx\left| v_\eta (x)\right| ^2\right] ,  \label{energy}
\end{eqnarray}
while the high-lying levels contribute 
\begin{equation}
E_{cont}=\int dx\left[ e_1\left( x\right) +e_2\left( x\right) \right] ,
\end{equation}
where $e_1\left( x\right) $ and $e_2\left( x\right) $ are given by, 
\begin{eqnarray}
e_1\left( x\right) &=&-\frac{\left( 2m\right) ^{1/2}}{4\pi \hbar }%
\int\limits_{E_c}^\infty \epsilon d\epsilon \left[ \frac{\epsilon +\delta
\mu }{\sqrt{\left( \epsilon +\delta \mu \right) ^2-\Delta ^2\left( x\right) }%
}-1\right]  \nonumber \\
&&\times \frac 1{\left[ \mu -V_{ext}\left( x\right) +\sqrt{\left( \epsilon
+\delta \mu \right) ^2-\Delta ^2\left( x\right) }\right] ^{1/2}},
\end{eqnarray}
and 
\begin{eqnarray}
e_2\left( x\right) &=&-\frac{\left( 2m\right) ^{1/2}}{4\pi \hbar }%
\int\limits_{E_c}^\infty \epsilon d\epsilon \left[ \frac{\epsilon -\delta
\mu }{\sqrt{\left( \epsilon -\delta \mu \right) ^2-\Delta ^2\left( x\right) }%
}-1\right]  \nonumber \\
&&\times \frac 1{\left[ \mu -V_{ext}\left( x\right) +\sqrt{\left( \epsilon
-\delta \mu \right) ^2-\Delta ^2\left( x\right) }\right] ^{1/2}},
\end{eqnarray}
respectively. An important self-consistency check of the calculations of the
entropy and of the energy is given by the thermodynamic relation, 
\begin{equation}
E\left( T\right) =E_0+\int_0^TdT^{\prime }T^{\prime }\left( \frac{\partial S}{\partial T^{\prime }}\right) ,
\end{equation}
where $E_0$ is the ground state energy at zero temperature.

Before closing the section, we discuss briefly the applicablity of 
mean-field Bogoliubov-de Gennes theory in one dimension. It is well-known 
that with decreasing dimensionality, the pair fluctuation becomes 
increasingly important. Specifically to 1D, the true long-range order 
is completely destroyed by fluctuations. Thus, strictly speaking, there is no \emph{a priori} 
justification for the use of mean-field BdG theory for a uniform spin-polarized 
Fermi gases. Nevertheless, we found in the previous study that at zero 
temperature and at a coupling constant $\gamma \sim 1$, the mean-field
BdG calculation provides reasonable description for the energy and chemical potential 
of a uniform spin-polarized Fermi gas, as compared to the exact Bethe Ansatz 
solutions (See, for example, the Figs. 12 and 13 in Ref. \cite{xiajipra1d}). 

On the other hand, the presence of a harmonic trap effectively increases 
the dimensionality of the system. For instance, the density of state of 
the trapped 1D system is a constant, exhibiting the same behavior as 
a uniform Fermi gas in two dimensions. Thus, in a harmonic trap, we antipicate 
that the pair fluctuations should be much suppressed, and thus the use of the mean-field 
BdG theory may be justified. We have checked in the previous work the 
validity of BdG theory for a trapped spin-polarized Fermi gas and found 
that indeed, at $\gamma \sim 1$ the density profile of a trapped gas obtained 
from the zero temperature BdG calculations agrees well with that from the 
asymptotically exact Gaudin solutions (See, for example, Fig. 18 in Ref. \cite{xiajipra1d}).

\section{Results and discussions}

For concreteness, we consider a trapped polarized gas with a number of total
atoms $N=N_{\uparrow }+N_{\downarrow }=100$. In traps,
it is convenient to take trap units with, $m=\hbar =\omega =k_B=1$, so that
the length and energy will be measured in units of the harmonic oscillator
length $a_{ho}=\sqrt{\hbar /m\omega }$ and of the level spacing 
$\hbar \omega $, respectively. The characteristic energy scale may be given
by the Fermi energy of an unpolarized ideal gas at zero temperature, 
$E_F=N\hbar \omega /2$, which provides also a characteristic temperature
scale, $T_F=E_F/k_B$. On the other hand, the length scale is set by a
Thomas-Fermi radius, $x_{TF}=N^{1/2}a_{ho}$. To characterize the
interaction, we use a dimensionless coupling constant at the trap center 
\cite{xiajipra1d}, $\gamma _0=a_{ho}/(\pi N^{1/2}a_{1D})$, where 
$a_{1D}=-mg_{1D}/(n\hbar ^2)$ is the 1D scattering length, and $\gamma _0$ is
roughly the ratio of the interaction energy density at the trap center to
the kinetic energy density. Thus, $\gamma_0 \ll 1$ corresponds to the weakly
interacting limit, while the strong coupling regime is realized when 
$\gamma_0 \gg 1$. Throughout the paper, we use a coupling constant 
$\gamma _0=1.6$, comparable to the estimates for a real experimental 
setup \cite{xiajipra1d}. 

%
\begin{figure}
\begin{centering}\includegraphics[clip,width=0.45\textwidth]{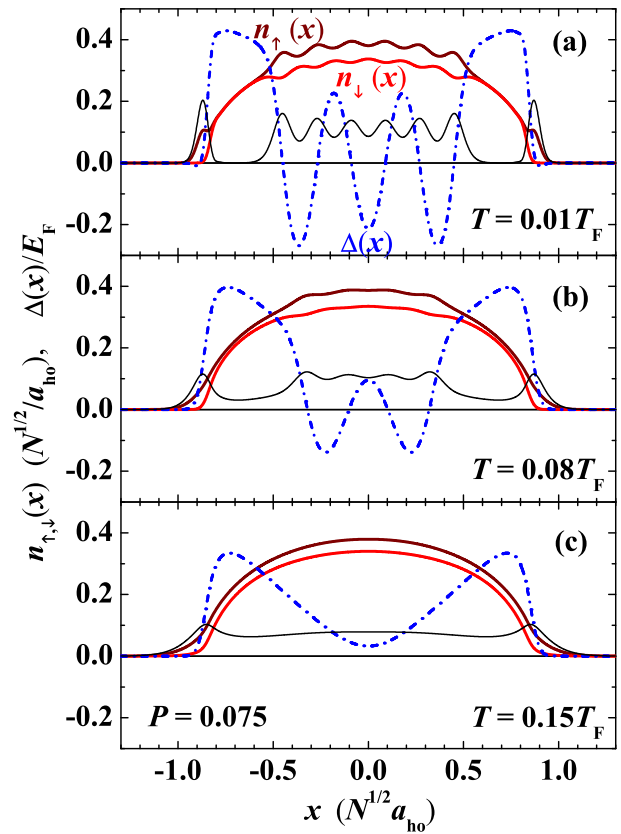}\par\end{centering}

\caption{(color online) Spin up and spin down density profiles (thick
solid lines), the density difference (thin solid lines), and the order
parameter (dot-dashed lines) of a trapped 1D Fermi gas for a spin
polarization $P=0.075$, at several temperatures, $T=0.01T_F$ (upper panel), 
$T=0.08T_F$ (middle panel), and $T=0.15T_F$ (bottom panel). For a better
illustration, the scale of the density difference has been doubled. At zero
temperature, the system stays in a phase separation phase with a FFLO state
at center and an outside BCS shell. The FFLO phase at the trap center,
signaled by the oscillations in the order parameter, suppresses with
increasing temperature, and leaves out a pure BCS state throughout the trap
at about $0.10T_F$. At a higher temperature around $0.20T_F$, the system
becomes fully normal through a second order superfluid phase transition.}

\label{fig2} 
\end{figure}


%
\begin{figure}
\begin{centering}\includegraphics[clip,width=0.45\textwidth]{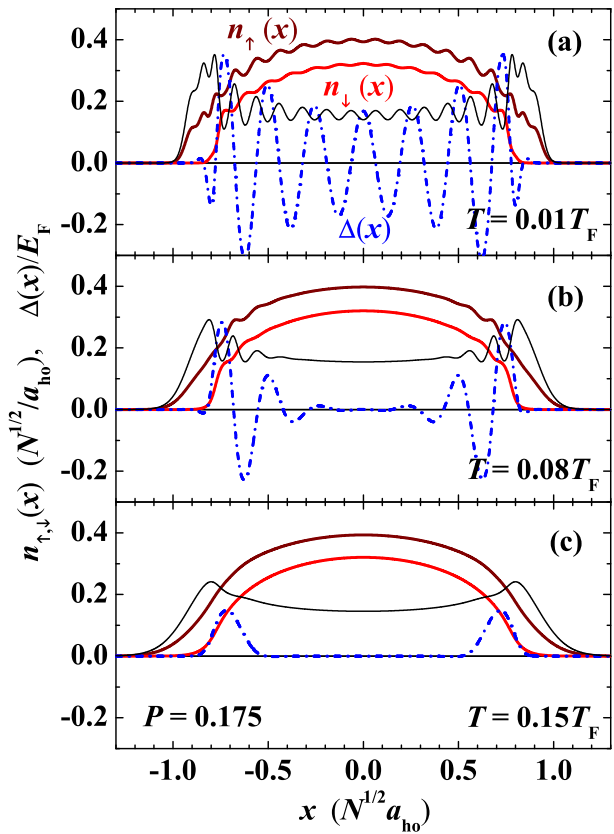}\par\end{centering}

\caption{(color online) Same plots as in Fig. 2, but for a larger spin
polarization $P=0.175$. With this value of spin polarizations, at zero
temperature the atoms at the trap edge stay in a normal state. With
increasing temperature, the FFLO phase at the trap center vanishes
gradually, which eventually leads to a normal phase throughout the trap at 
$T\simeq 0.15T_F$.}

\label{fig3} 
\end{figure}


\subsection{Density profiles and order parameter}

Figs. 2 and 3 give BdG results for the density profiles of each
component, as well as the density difference $\delta n\left( x\right)
=n_{\uparrow }\left( x\right) -n_{\downarrow }\left( x\right) $ and the
order parameter $\Delta (x)$, at different temperatures as indicated and at
two spin polarizations, $P=(N_{\uparrow }-N_{\downarrow })/N=0.075$ (Fig. 2)
and $0.175$ (Fig. 3). In the work by Orso \cite{orso} and our previous 
studies \cite{xiajiprl,xiajipra1d}, it has shown that at zero temperature 
there are two phase separation states: For a small spin polarization the 
system stays in the spatially inhomogeneous 
FFLO superfluid state at the trap center and in a BCS superfluid state at 
the edge (referred to as FFLO-BCS later). For a large spin polarization, 
the gas still remains in the FFLO state at the trap center, but becomes a 
fully polarized normal state towards the edge of the trap (referred to as 
FFLO-N later). We refer to the Secs. VI and VIIA of Ref. \cite{xiajipra1d} 
for further details. The two spin polarizations in Figs. 2 and 3 are selected 
in such a way that initially (at low temperature) the cloud is in 
different phase separation ground state. We then trace how these two 
phase separation states develop as the temperature increases.

For a small spin polarization below a critical value $P_c$, the FFLO-BCS
phase separation state at low temperature is clearly characterized by the
spatial profile of the order parameter (Fig. 2a). It shows oscillations at
the trap center, characteristic of FFLO states, and two shoulders at the trap
edge, characteristic of a 1D BCS state. Naively, the FFLO state is much more easily
disturbed by thermal fluctuations than the BCS state, due to a
reduced pairing phase space. Therefore, as shown in Figs. 2b and 2c, with
increasing temperature the FFLO superfluid is suppressed faster than
the BCS shoulders at the edge. This leads to a pure BCS state above a
certain temperature about $0.10T_F$. Further increase of the temperature
will destroy the superfluid state eventually at around $0.20T_F$.

%
\begin{figure}
\begin{centering}\includegraphics[clip,width=0.45\textwidth]{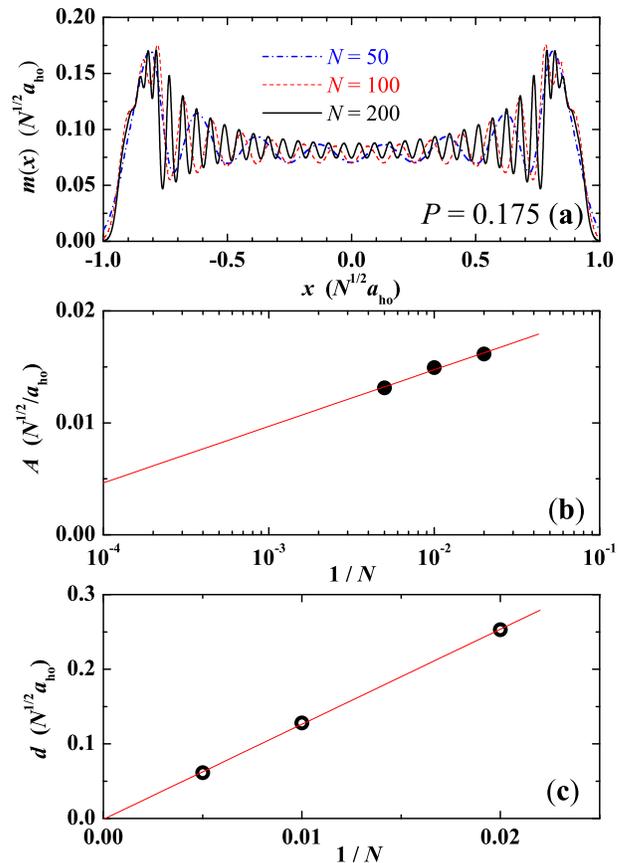}\par\end{centering}

\caption{(color online) (a) Density difference of a spin-polarized Fermi gas
at $P=0.175$ and $T=0.01T_F$ for different numbers of total atoms as indicated. 
(b) The number dependence of the amplitude of the density oscillations $A$ at 
the trap center. The amplitude $A$ decreases slowly with increasing the number of 
atoms. (c) The number dependence of the periodicity of the density oscillations 
$d$ at the trap center.}

\label{fig4} 
\end{figure}


%
\begin{figure}
\begin{centering}\includegraphics[clip,width=0.45\textwidth]{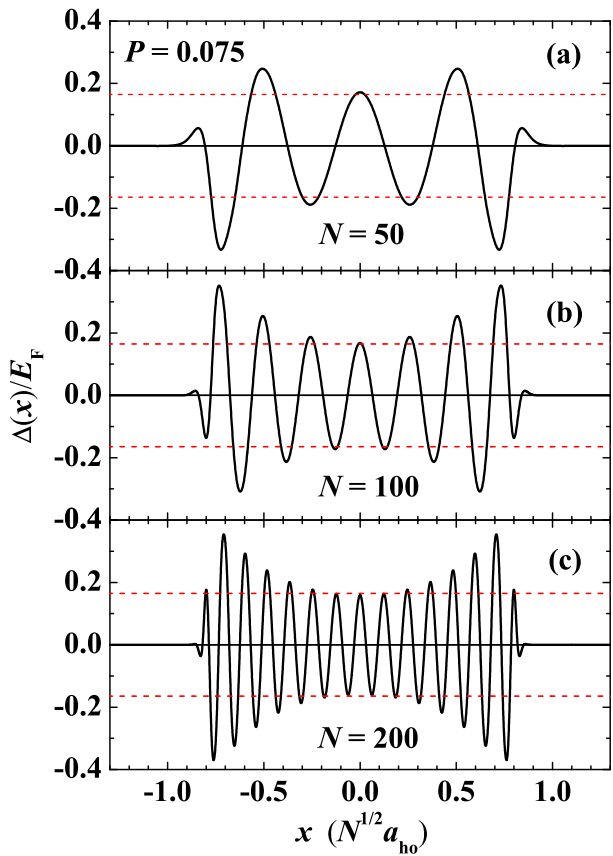}\par\end{centering}

\caption{(color online) Order parameters of a spin-polarized Fermi gas
at $P=0.175$ and $T=0.01T_F$ for different numbers of total atoms: 
(a) $N=50$, (b) $N=100$, and (c) $N=200$. The two dashlines indicate the value of the
order parameter at the trap center.}

\label{fig5} 
\end{figure}


Experimentally, the change from FFLO-BCS phase separation state to the pure
BCS state and finally to the normal state may be monitored by measuring the
density difference. At low temperature in FFLO-BCS separation phase, the
local spin polarization or density difference is fully carried by the FFLO
state and thus is restricted to the trap center (Fig. 2a). Here, we have 
neglected the single peak of density difference exactly at the trap edge 
that is caused by breakdown of the mean-field approach, 
see Ref. \cite{xiajipra1d} for detailed discussions. At a higher temperature, 
the BCS state starts to contribute to the spin polarization, due to the 
thermal excitations. As a result, the density difference leaks out 
gradually to the trap edge (Fig. 2b). When the FFLO state fully disappears, 
the difference in density becomes very flat throughout the trap (Fig. 2c).

The temperature evolution for a FFLO-N phase separation state with large
spin polarizations is simpler, as shown in Fig. 3. In this case, the FFLO state at the trap
center is destroyed gradually by increasing temperature, and one ends up with a
fully normal cloud around $T\simeq 0.15T_F$. The shape of density difference
profiles is nearly independent of temperature. Thus, the distinct temperature
dependence of the density difference in the FFLO-BCS and FFLO-N states
provides a useful way to distinguish these two phase separation phases.

At the end of this subsection, we emphasize that in Figs. (2a) and (2b) it
is the oscillation of the order parameter that is the unambiguous signature
of the FFLO state. There is also a related oscillation in the density
profiles with doubling periodicity. However, these density oscillations are
presumably due to finite size effect. We have checked this point by varying
the number of total atoms. In Figs. 4 and 5, we show the density difference
and the order parameter at a spin polarization $P=0.175$ and at a low
temperature $T=0.01T_F$ for particle numbers in the range $50\sim 200$. With
increasing the number of total atoms, the amplitude of the density
oscillation at the trap center becomes smaller. In contrast, the amplitude
of the order parameter $\Delta \left( x=0\right) $ stays nearly constant.
Note that in Fig. 5 we find a Larkin-Ovchinnikov (LO) state in the center region, 
\textit{i.e.}, the order parameter behaves like $\Delta \left( x\right) =\Delta
\left( x=0\right) \cos (q_{\text{LO}}x)$ with a center-of-mass momentum $q_{\text{LO}}$
proportional to the Fermi wave-vector difference, $q_{\text{LO}}\simeq
k_{F\uparrow }-k_{F\downarrow }\simeq (N^{1/2}P)a_{ho}^{-1}$. Thus, in units
of $(N^{1/2}a_{ho})$ the periodicity of the density oscillation (Fig. 4c)
and of the order parameter are inversely proportional to the number of total atoms.

%
\begin{figure}
\begin{centering}\includegraphics[clip,width=0.45\textwidth]{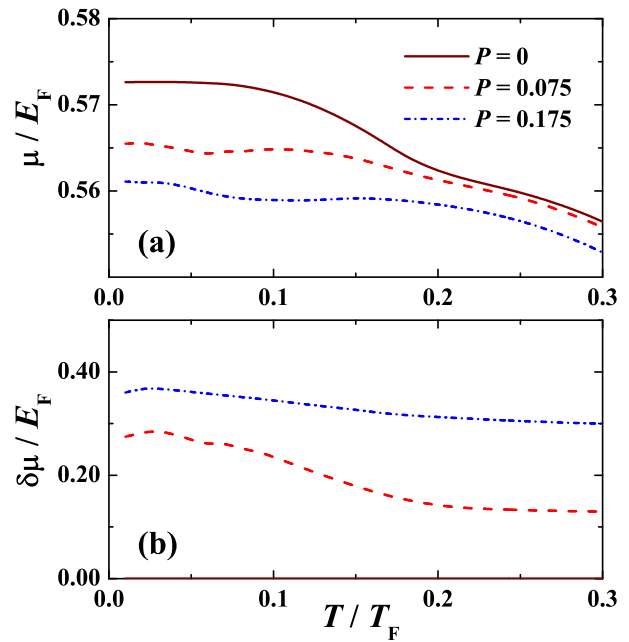}\par\end{centering}

\caption{(color online) Temperature dependence of the chemical
potential (panel a) and of the chemical potential difference (panel b) at
three spin polarizations: $P=0$ (solid lines), $P=0.075$ (dashed lines), and 
$P=0.175$ (dash-dotted lines).}

\label{fig6} 
\end{figure}


%
\begin{figure}
\begin{centering}\includegraphics[clip,width=0.45\textwidth]{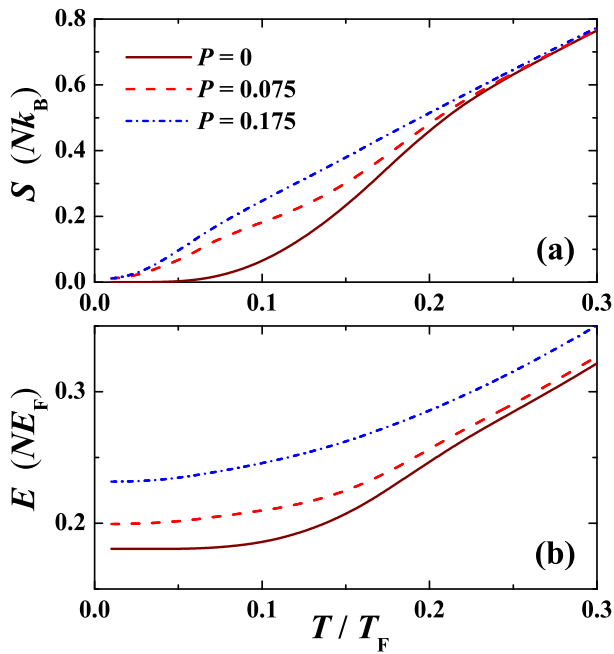}\par\end{centering}

\caption{(color online) Temperature dependence of the total entropy per
particle (panel a) and of the total energy per particle (panel b) at three
spin polarizations: $P=0$ (solid lines), $P=0.075$ (dashed lines), 
and $P=0.175$ (dash-dotted lines).}

\label{fig7} 
\end{figure}


\subsection{Equation of state}

Figs. 6 and 7 show the equation of state as a function of temperature. The
total entropy and total energy increase monotonically with increasing
temperature and spin polarization. In particular, at a finite spin
polarization, the entropy increases linearly with temperature, as compared to
the exponential temperature dependence for a BCS superfluid (\textit{i.e.}, 
$P=0$). This linear dependence arises from having gapless single-particle
excitations located at the node of FFLO states at the trap center, and in
case of the FFLO-N phase, from the normal cloud at the edge of the trap. 
As the temperature increases, nontrivial kinks are visible in both
chemical potentials and entropy, reflecting the smooth transitions between
different states as mentioned earlier.

%
\begin{figure}
\begin{centering}\includegraphics[clip,width=0.45\textwidth]{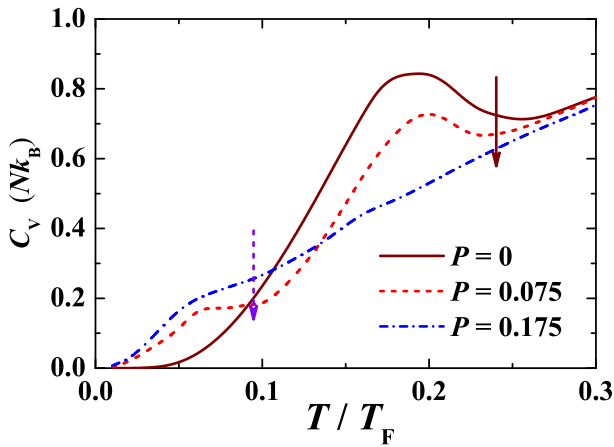}\par\end{centering}

\caption{(color online) Temperature dependence of the specific heat at
three spin polarizations: $P=0$ (solid lines), $P=0.075$ (dashed lines), and 
$P=0.175$ (dash-dotted lines). For $P=0.075$, the dashed arrow indicates the
transition from a FFLO-BCS phase separation phase to a pure BCS state, while
the solid arrow marks the second order phase transition from a BCS
superfluid to a normal gas.}

\label{fig8} 
\end{figure}


\subsection{Specific heat and finite temperature phase diagram}

The analysis of the profiles of the density distributions and of the order
parameter has already given some qualitative features for the phase diagram
at finite temperatures. To quantify it, we calculate the specific heat,
using the definition,
\begin{equation}
C_V=\frac{\partial E}{\partial T}=T\frac{\partial S}{\partial T}.
\end{equation}
Fig. 8 displays the temperature dependence of the specific heat at three
spin polarizations. Different transitions are well characterized by the
apparent kinks in the specific heat. For a small polarization, two kinks are
clearly visible, corresponding to the transition from a FFLO-BCS phase
separation state to a pure BCS state, and in turn to a normal state. 
In contrast, for a large polarization, typically only one kink is identifiable, 
which should be attributed to the transition from a FFLO-N phase 
separation state to a completely normal state.

Gathering the position (temperature) of the kinks for different spin
polarizations, we obtain the solid and dashed lines in the phase diagram, as
shown in Fig. 1. The former line corresponds to the second-order phase transition
from either a BCS or FFLO-N to the normal state; while the later distinguishes
a high-$T$ pure BCS state from a low-$T$ FFLO-BCS phase separation state. The 
determination of the boundary between FFLO-BCS and FFLO-N phase separation 
phases, for example, the critical spin polarization $P_c(T)$, is more difficult 
from numerics. As discussed in our previous works \cite{xiajiprl,xiajipra1d}, 
the critical spin polarization is calculated from a critical chemical potential 
difference, \textit{i.e.}, the half of binding energy of the 1D molecule pairs. 
From Fig. 6b, for a large spin polarization the chemical potential difference 
is temperature insensitive. On quite general grounds, we thus assume that 
the critical spin polarization $P_c(T)$ is nearly temperature independent. 
This gives the dash-dotted line in the phase diagram (Fig. 1).

It is evident from Fig. 1 that there is a wide temperature window for the
presence of FFLO states at the trap center. The typical temperature for
observing the FFLO state would be of one-tenth of the Fermi temperature,
corresponding to an entropy $S\sim 0.2Nk_B$. Such a temperature or entropy
is within the reach of present-day techniques \cite{natphys}.

\section{Concluding remarks}

In conclusion, based on a mean-field Bogoliubov-de Gennes theory, we
calculate the thermodynamic properties of a polarized atomic Fermi gas in a
highly elongated harmonic trap. The profiles of the density distributions and of
order parameter, the equation of state, as well as the specific heat have
been analyzed in detail. We have then established a finite temperature phase
diagram. Our results are useful for experiments at Rice 
University \cite{hulet}, in which a search for the exotic FFLO states 
in 1D polarized Fermi gases is under-taken. Our estimated temperature and 
entropy for the realization of FFLO states are about $0.1T_F$ 
and $0.2Nk_B$, respectively. These values are experimentally 
attainable \cite{natphys}.

While our weak-coupling Bogoliubov-de Gennes study provides a
semi-quantitative finite-temperature phase diagram of the 1D polarized Fermi
gas, an improved description could be obtained by solving the exact
thermodynamic Bethe ansatz solutions of the 1D gas \cite{takahashi}, 
with the trap effect treated using a local density 
approximation \cite{orso,xiajiprl}. We plan on doing this in future work.

\begin{acknowledgments}
This research is supported by the Australian
Research Council Center of Excellence, the National Natural Science
Foundation of China Grant No. NSFC-10774190, and the National Fundamental
Research Program of China Grant Nos. 2006CB921404 and 2006CB921306. 
\end{acknowledgments}

\end{document}